\documentclass[journal]{IEEEtran}
\IEEEoverridecommandlockouts
\usepackage{cite}
\usepackage{breakurl}
\usepackage{amsmath,amssymb,amsfonts}
\usepackage{algorithmic}
\usepackage{graphicx}
\usepackage{textcomp}
\usepackage{xcolor}
\usepackage{subcaption}
\usepackage{empheq}
\def\BibTeX{{\rm B\kern-.05em{\sc i\kern-.025em b}\kern-.08em
		T\kern-.1667em\lower.7ex\hbox{E}\kern-.125emX}}

\usepackage{bbding}
\usepackage{pifont}
\usepackage{wasysym}
\usepackage{amssymb}

\usepackage[justification=centering]{caption}

\usepackage{diagbox}
\usepackage{mathtools}
\usepackage[left=0.65in,top=0.76in,right=0.65in,bottom=0.76in]{geometry}
\makeatletter
\let\old@ps@headings\ps@headings
\let\old@ps@IEEEtitlepagestyle\ps@IEEEtitlepagestyle
\def\confheader#1{
	\def\ps@IEEEtitlepagestyle{
		\old@ps@IEEEtitlepagestyle
		\def\@oddhead{\normalfont\footnotesize\centering#1}%
	}%
}

\makeatother

\confheader{%
	This paper has been accepted for publication in IEEE International Conference on Communications Workshops (ICC Workshops), 20-24 May 2019, Shanghai, China
}

\begin{document}

\title{An Efficient Blockchain-based Hierarchical Authentication Mechanism for Energy Trading in V2G Environment}

\author{
	\IEEEauthorblockN{Sahil Garg\IEEEauthorrefmark{1}, Member, IEEE, Kuljeet Kaur\IEEEauthorrefmark{1}, Member, IEEE, Georges Kaddoum\IEEEauthorrefmark{1}, Member, IEEE, \\Fran\c{c}ois Gagnon\IEEEauthorrefmark{1}, Senior Member, IEEE, and Joel~J.~P.~C.~Rodrigues\IEEEauthorrefmark{2}, Senior Member, IEEE}\\
	\IEEEauthorblockA{\IEEEauthorrefmark{1}Electrical Engineering Department, \'Ecole de technologie sup\'erieure, Montr\'eal, QC H3C 1K3, Canada.\\
		\IEEEauthorrefmark{2}National Institute of Telecommunications (Inatel), Santa Rita do Sapuca\'{i}, MG, Brazil; Instituto de Telecomunica\c{c}\~oes, Portugal; Federal University of Piau\'i (UFPI), Teresina, PI, Brazil.\\}
	E-mail: sahil.garg@ieee.org, kuljeet.kaur@ieee.org, georges.kaddoum@etsmtl.ca, \\francois.gagnon@etsmtl.ca, and joeljr@ieee.org
}

\maketitle
\begin{abstract}
	Vehicle-to-grid (V2G) networks have emerged as a new technology in modern electric power transmission networks. It allows bi-directional flow of communication and electricity between electric vehicles (EVs) and the Smart Grid (SG), in order to provide more sophisticated energy trading. However, due to the involvement of a huge amount of trading data and the presence of untrusted entities in the visiting networks, the underlying V2G infrastructure  suffers from various security and privacy challenges. Although, several solutions have been proposed in the literature to address these problems, issues like lack of mutual authentication and anonymity, incapability to protect against several attack vectors, generation of huge overhead, and dependency on centralized infrastructures make security and privacy issues even more challenging. To address the above mentioned problems, in this paper, we propose a blockchain oriented hierarchical authentication mechanism for rewarding EVs. The overall process is broadly classified into the following phases: 1) System Initialization, 2) Registration, 3) Hierarchical Mutual Authentication, and 4) Consensus; wherein blockchain's distributed ledger has been employed for transaction execution in distributed V2G environments while Elliptic curve cryptography (ECC) has been used for hierarchical authentication. The designed hierarchical authentication mechanism has been employed to preserve the anonymity of EVs and support mutual authentication between EVs, charging stations (CSs) and the central aggregator (CAG). Additionally, it also supports minimal communicational and computational overheads on resource constrained EVs. Further, formal security verification of the proposed scheme on widely accepted Automated Validation of Internet Security Protocols and Applications (AVISPA) tool validates its safeness against different security attacks.
\end{abstract}
	
	
\section{Introduction}
With the rapid proliferation of Information and Communication Technologies (ICT), smart grids (SGs) are gaining tremendous attention. It uses bi-directional flow of information and electrical energy to create an intelligent and widely distributed automated energy networks. As an important component, Vehicle-to-Grid (V2G) networks have emerged, wherein electric vehicles (EVs) interacts with SGs, especially for energy trading. Here, EVs with surplus energy perform charge and discharge operations in order to balance the power demand of SGs \cite{Kaur:2018:GIE:3243318.3243322}. However, due to the association of vehicle mobility, charging and discharging operations, and limited communication range, the information shared across EVs and other V2G entities face significant security and privacy risks. Thus, EVs, which play a key role in energy transportation and management, may not be willing to participate in energy trading. In order to encourage EVs in energy trade-off, it is prevalent to design a secure, efficient, and reliable authentication mechanism for energy trading in V2G setups. \\
\indent Recently, several cryptographic schemes based on authentication, physical layer protection, and encryption have been proposed for SGs. For example, Odelu \textit{et al.} \cite{7549086} proposed a secure authenticated key agreement scheme for SG, which provides privacy and session-key security under the Canetti-Krawczyk adversary model. Eiza \textit{et al.} \cite{8531779} designed an efficient, secure and privacy-preserving proxy mobile IPv6 (PMIPv6) protocol to address the security and privacy concerns of mobile IP communications in V2G networks. In a similar direction, Wu \textit{et al.} \cite{8519740} utilized elliptic curve cryptography (ECC) to propose a secure and lightweight agreement mechanism for SG. In order to assure the confidentiality and integrity of V2G connections, Abdallah and Shen \cite{7485857} designed a lightweight authentication and privacy-preserving scheme where EVs are allowed to generate their own pseudonym identities for protecting their private information. Likewise, Kumar \textit{et al.} \cite{8413131} designed a hybrid cryptography based authentication and key agreement scheme to facilitate mutual trust between the legitimate entities in smart energy networks. In order to provision the authentication between EVs and smart meters, Wazid \textit{et al.} \cite{7995139} devised a three-factor user authentication scheme for SG environments based on lightweight cryptographic primitives such as one-way hash functions, bitwise XOR operations and ECC. Similarly, Gope and Sikdar \cite{8373734} used physically uncloneable functions and one-way hash functions to develop a privacy-aware authenticated key agreement scheme for SG communications. In order to address the security and privacy issues in the V2G networks, Shen \textit{et al.} \cite{8115145} proposed a robust key agreement protocol by leveraging hash functions and bitwise exclusive-OR operations. \\
\indent Although several authentication schemes have been proposed, most of them are deemed unfit for resource constrained V2G setups since they depend on public-key cryptosystems, cannot ensure security against insider attacks, lacks anonymity, incur high communication and computation costs, and moreover suffer from problems like single point of failure and privacy leakage \cite{7841955}. In order to support an adequate level of security in V2G setups, a promising blockchain technology has been introduced because of its high potential to support decentralization, anonymity, trust, and integrity, with a moderate cost. It is a peer-to-peer (P2P) distributed ledger technology that maintains transactional data across several systems in a verifiable and permanent manner. It adopts multiple means such as data encryption, automated scripts, distributed consensus, time stamping, and economic incentives in order to improve the security, intelligence, storage, and management while solving the problems of high costs and inefficiency, that are common in traditional centralized energy trading systems. \\
\indent In this direction, several authors used blockchain for solving the security and privacy concerns in decentralized SG environments. For example, Guan \textit{et al.} \cite{8419184} proposed a blockchain based privacy-preserving and data aggregation scheme for secure communications in SG, where Bloom filter was adopted to realize the fast authentication. Likewise, Wang \textit{et al.} \cite{8598651} designed an efficient anonymous rewarding scheme which employs digital signature, ring signature, encryption, blockchain, and Monero to satisfy the security requirements of V2G networks. In a similar direction, Liu \textit{et al.} \cite{8582227} devised a cross-domain authentication scheme, where consortium blockchain and SM9, an identity-based cryptographic algorithm, was employed to provide the required level of security and privacy in V2G networks. In order to address the security and privacy challenges caused by untrusted parties, Li \textit{et al.} \cite{8234700} proposed a consortium blockchain based solution for secure energy trading in Industrial Internet of Things (IIoT). Moreover, they also proposed a credit-based payment scheme to support a fast and frequent P2P trading of energy; wherein Stackelberg game was used to decide the optimal pricing strategy. Similarly, Kang \textit{et al.} \cite{7935397} also deployed a consortium blockchain technology to address the security and privacy issues in P2P energy trading among EVs. Likewise, Aitzhan and Svetinovic \cite{7589035} also addressed the security and privacy issues of energy trading data using blockchain, multi-signatures, and anonymous message propagation streams. Although several schemes have been proposed in the literature, they may not work well due to the lack of mutual authentication between a communicating parties and inability to preserve their anonymity. Thus, in this paper we employ a combination of ECC and blockchain for secure and anonymous energy trading in V2G setups. 

\subsection{Contributions} \label{sec:DesignGoals}
\noindent Key contributions of this research work are illustrated below:

\begin{itemize}
	\item We present an effective blockchain based hierarchical authentication mechanism for secure and anonymous energy trading in V2G setups. Here, blockchain's distributed ledger is employed for transaction execution in distributed V2G environments while ECC is used for hierarchical authentication.
	\item The hierarchical authentication mechanism has been designed to preserve the anonymity of EVs and support mutual authentication between the EVs, charging stations (CSs) and central aggregator (CAG). Additionally, it also supports minimal communicational and computational overheads on resource constrained EVs.
	\item We also justify the performance of the proposed scheme on the widely acceptable AVISPA tool and establish the reduced burden on EVs for participating in secure V2G energy trading mechanism. 
\end{itemize}

\subsection{Organization}
The rest of the manuscript is structured in accordance with the following sequence: Section~\ref{sec:SystemModel} presents the high level description of the proposed system model followed by the proposed scheme in Section~\ref{sec:ProposedScheme}. The detailed security analysis of the proposed scheme along with extensive performance assessment are presented in Section~\ref{sec:ObservationAndAnalysis}. Finally, Section~\ref{sec:Conclusions} concludes the proposed work.

\section{System Model} \label{sec:SystemModel}
This section illustrates the high level view of the considered V2G scenario with different entities helping in forming and maintaining the distributed  ledger. 

\begin{figure}[t]
	\centering
	\includegraphics[scale=.5]{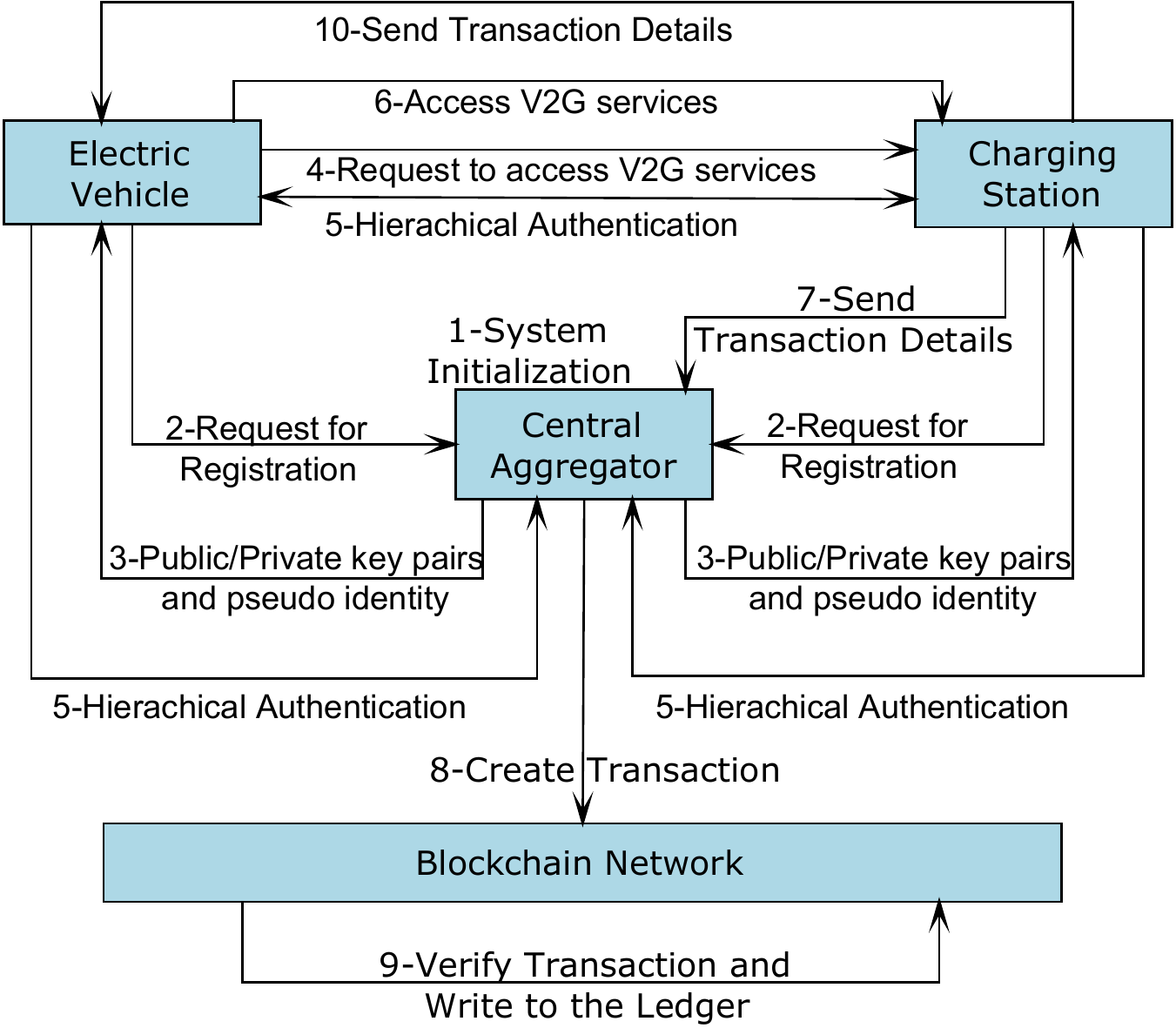}
	\caption{Systematic diagram of the proposed scheme.}
	\label{fig:SystemModel}
\end{figure}

Fig.~\ref{fig:SystemModel} depicts the systematic diagram of the proposed scheme with different components of the considered ecosystem and their corresponding execution steps. As evidenced from the figure, the considered setup is comprised of four core entities namely-EVs, CSs, CAG, and the blockchain network. The fleet of EVs are distributed energy sources that help in maintaining the SG's stability either by injecting or withdrawing energy from the grid. These V2G services help in stabilizing the SG's operations in both peak and off-peak hours. Hence, SG imparts incentives to the EVs for participating in the regulatory process. On the other hand the EVs' charge and discharge their respective batteries at dedicated charging points available at the CS level. These CSs are equipped with smart meters (to keep track of the amount of energy withdrawn/injected) and record the current electricity prices. Thus, CSs know the amount of rewards that needs to be paid to an EV for participating in the regulatory mechanism and is responsible for generating the related transactions. Above all, the CAG is the central authority that validates the transactions created by CSs and maintains the entire blockchain network (with the help of CSs). Additionally, it is also responsible for registering legitimate a nd illegitimate EVs and CSs. Here, the blockchain network helps in transmitting the rewards to designated EVs in a secure and anonymous manner.

The foremost step in the proposed V2G energy-trading process is the system initialization wherein CAG releases the public parameters for implementing the cryptographic functions. In the next step, EVs and CSs register themselves with the CAG and obtain their respective pseudo identity. These identities of EVs correspond to their address on the global ledger. Additionally, the CAG is also responsible for generating public-private key pairs for all EVs, CSs and itself. For secure energy trading, the CSs accept an EV's request to participate post successful mutual authentication between itself, the EV and the CAG. Once the authenticity is established, the CS provides charging/discharging services to the designated EV and generates the corresponding reward results. The results are transferred to the CAG which verifies the transaction and writes a block to the ledger using the practical byzantine fault tolerance (PBFT) mechanism. With consensus establishment, the reward is transferred to the designated EV and a receipt is sent to the EV by the CS. 

\section{Proposed Scheme} \label{sec:ProposedScheme}	
The overall process of blockchain based hierarchical authentication mechanism for rewarding EVs can be broadly classified into the following phases: 1) System Initialization, 2) Registration, 3) Hierarchical Mutual Authentication, and 4) Consensus. The detailed information about these phases is summarized as follows:

\vspace{2.5mm}
\noindent \textit{\textbf{Phase 1: System Initialization Phase}}
\vspace{1mm}

During this phase, the CAG prepares the V2G environment for subsequent phases as follows:

\vspace{0.7mm}
\noindent \textit{Step 1:} The CAG selects an elliptic curve $E$ with base point $P$ and a large prime number $n$.

\vspace{0.7mm}
\noindent \textit{Step 2:} Using the above parameters, the CAG generates its private key $\mathbb{SK}_{AD} \in  Z_p^{*}$. Following this, the CAG employ a ECC multiplicative operation over $\mathbb{SK}_{CAG}$ to generate its public key as follows: $\mathbb{PK}_{CAG}= \mathbb{SK}_{CAG}.P$.

\vspace{0.7mm}
\noindent \textit{Step 3:} The CAG publishes all the public parameters including: $E$,  $p$, $n$, $\mathbb{PK}_{CAG}$, and the one-way collision resistant hash function $\mathbb{H}(.)$.

\vspace{2.5mm}
\noindent \textit{\textbf{Phase 2:  Registration}}
\vspace{1mm}

\begin{figure}[h]
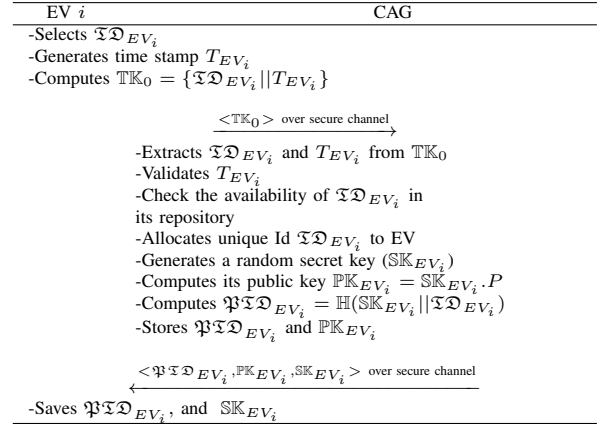

	\centering
	\scriptsize
	\begin{tabular}{p{1cm} p{.5cm} p{5cm}}
		\hline
		\multicolumn{1}{c}{{EV $i$}} & {}  & \multicolumn{1}{c}{{CAG}}   \\
		\hline
		\multicolumn{3}{l}{-Selects $\mathfrak{TD}_{EV_i}$}\\
		\multicolumn{3}{l}{-Generates time stamp $T_{EV_i}$}\\
		\multicolumn{3}{l}{-Computes $\mathbb{TK}_0=\{\mathfrak{TD}_{EV_i} || T_{EV_i}\}$}\\
		
		&&\\
		\multicolumn{3}{c}{$\xrightarrow{<\mathbb{TK}_0>~\text{over secure channel} }$}\\
		
		&\multicolumn{2}{l}{-Extracts  $\mathfrak{TD}_{EV_i}$  and $T_{EV_i}$ from $\mathbb{TK}_0$}\\
		&\multicolumn{2}{l}{-Validates $T_{EV_i}$}\\
		& \multicolumn{2}{l}{-Check the availability of $\mathfrak{TD}_{EV_i}$ in }\\
		& \multicolumn{2}{l}{its repository}\\
		&\multicolumn{2}{l}{-Allocates unique Id $\mathfrak{TD}_{EV_i}$ to EV}\\
		&\multicolumn{2}{l}{-Generates a random secret key ($\mathbb{SK}_{EV_i}$)}\\
		&\multicolumn{2}{l}{-Computes its public key $ \mathbb{PK}_{EV_i}=\mathbb{SK}_{EV_i}.P$}\\
		
		& \multicolumn{2}{l}{-Computes $\mathfrak{PTD}_{EV_i}=\mathbb{H}(\mathbb{SK}_{EV_i}|| \mathfrak{TD}_{EV_i})$}\\
		&\multicolumn{2}{l}{-Stores $\mathfrak{PTD}_{EV_i}$ and $\mathbb{PK}_{EV_i}$}\\
		
		&&\\
		\multicolumn{3}{c}{$\xleftarrow{<\mathfrak{PTD}_{EV_i}, \mathbb{PK}_{EV_i}, \mathbb{SK}_{EV_i}>~\text{over secure channel}}$}\\
		\multicolumn{3}{l}{-Saves $\mathfrak{PTD}_{EV_i},$ and $~\mathbb{SK}_{EV_i}$}\\
		
		\hline
	\end{tabular}
	\caption{An illustration of the EV's registration process.}
	\label{fig:Phase2}
\end{figure}

This phase involves the registration of the legitimate EVs and CSs at the CAG level over a secure channel. The process of registering an EV and a CS is identical. Hence, this sub-section elaborates the registration process for the $i^{th}$ EV. The pictorial representation of this process can be found in Fig.~\ref{fig:Phase2}.

\vspace{0.7mm}
\noindent \textit{Step 1:} EV $i$ selects an identity $\mathfrak{TD}_{EV_i}$ for uniquely presenting itself. This identity could be EV's license number or it vehicle identification number issued by the auto-mobile company. Next, the EV generates the current time stamp $T_{EV_i}$ and computes the token $\mathbb{TK}_0=\{\mathfrak{TD}_{EV_i} || T_{EV_i}\}$.

\vspace{0.7mm}
\noindent \textit{Step 2:} The value of the token $\mathbb{TK}_0$ is then transmitted to the CAG over the secure channel.
	
\vspace{0.7mm}
\noindent \textit{Step 3:} Upon receiving the token value, the CAG extracts $\mathfrak{TD}_{EV_i}$ and $T_{EV_i}$. Following this, it validates the time stamp and proceeds further only if it is within the permissible range.

\vspace{0.7mm}
\noindent \textit{Step 4:} The CAG verifies the existence of $\mathfrak{TD}_{EV_i}$ in its repository and revocation list. A match found in its repository indicates that the $i^{th}$ EV has been registered earlier. On the other hand, a match in the revocation list denotes that the EV is illegitimate and should not be registered. In either cases, the connection is terminated.

\vspace{0.7mm}
\noindent \textit{Step 5:} In this step, the CAG accepts the EV's request for registration and generates a public-private key pair ($\mathbb{PK}_{EV_i}$ \& $\mathbb{SK}_{EV_i}$) for it using ECC.

\vspace{0.7mm}
\noindent \textit{Step 6:} The CAG also computes a pseudo identity for the $i^{th}$ EV as follows: $\mathfrak{PTD}_{EV_i}=\mathbb{H}(\mathbb{SK}_{EV_i} || \mathfrak{TD}_{EV_i}$. Finally, the computed keys and pseudo identity are trasmitted to the EV over the secure channel. 

\vspace{0.7mm}
\noindent \textit{Step 7:} The CAG stores the $\mathfrak{PTD}_{EV_i}$ and $\mathbb{PK}_{EV_i}$ values; while EV stores the $\mathfrak{PTD}_{EV_i}$ and $\mathbb{SK}_{EV_i}$ values.

\vspace{2.5mm}
\noindent \textit{\textbf{Phase 3: Hierarchical Mutual Authentication}}
\vspace{1mm}

During this phase, the $i^{th}$ EV mutually authenticates the CAG using the $j^{th}$ CS before commencing any transaction. The proposed authentication mechanism relies on inexpensive ECC, one-way hash functions, and concatenation operations, and is referred to as the hierarchical authentication mechanism. The detailed execution process is illustrated in Fig.~\ref{fig:Phase3} and described in detail as follows: 
\begin{figure*}[t!h]
	\centering
		\scriptsize
	\begin{tabular}{p{2cm}p{2cm}p{4cm}}
		\hline
		\multicolumn{1}{l}{{EV $i$}} & \multicolumn{1}{l}{CS $j$}  & \multicolumn{1}{l}{{CAG}}   \\
		\hline
		
		&\multicolumn{2}{l}{-Select a random number $r_1\in Z^{*}_p$ }\\
		&\multicolumn{2}{l}{-Generate time stamp $T_{CS_j}$ }\\
		&\multicolumn{2}{l}{-Compute $R_1=r_1.P$ }\\
		&\multicolumn{2}{l}{-Compute $M_1= <R_1 || \mathfrak{PTD}_{CS_j} || T_{CS_j}>$ }\\
		\multicolumn{2}{l}{$\xleftarrow{M_1= <R_1 || \mathfrak{PTD}_{CS_j} || T_{CS_j}>}$} &\\
		
		\multicolumn{3}{l}{-Extract $R_1$, $\mathfrak{PTD}_{CS_j}$ and $T_{CS_j}$}\\ 
		\multicolumn{3}{l}{Validate $T_{CS_j}$}\\
		\multicolumn{3}{l}{-Compute $R_2=\mathbb{SK}_{EV_i}. R_1$ }\\
		\multicolumn{3}{l}{-Generate time stamp $T_{EV_i}$ }\\
		\multicolumn{3}{l}{-Compute $\mathbb{A}uth_{EV_i-CS_j}= \mathbb{H}(R_1 || R_2 || \mathfrak{PTD}_{EV_i} || \mathfrak{PTD}_{CS_j} || T_{EV_i} ) $}\\
		\multicolumn{3}{l}{-Compute $\mathbb{A}uth_{EV_i-CAG}= \mathbb{H}(R_1 || R_2 || \mathfrak{PTD}_{EV_i} || \mathfrak{PTD}_{CAG} || T_{EV_i} ) $}\\
		
		\multicolumn{3}{l}{$\xrightarrow{M2=<\mathbb{A}uth_{EV_i-CS_j},~\mathbb{A}uth_{EV_i-CAG},~T_{EV_i}, ~\mathfrak{PTD}_{EV_i}>}$}\\
		
		&\multicolumn{2}{l}{-Validate time stamp $T_{EV_i}$}\\
		&\multicolumn{2}{l}{-Compute $\mathbb{A}uth_{EV_i-CS_j}^{*}= \mathbb{H}(R_1 || r_1.\mathbb{PK}_{EV_i}|| \mathfrak{PTD}_{EV_i} || \mathfrak{PTD}_{CS_j} || T_{EV_i} ) $}\\
		&\multicolumn{2}{l}{-Check $\mathbb{A}uth_{EV_i-CS_j}^{*} \stackrel{?}{=}  \mathbb{A}uth_{EV_i-CS_j}$}\\
		&\multicolumn{2}{l}{-If same, EV $i$ is marked authentic; else tear down the connection}\\
		
		&\multicolumn{2}{l}{-Select a random number $r_2\in Z^{*}_p$} \\
		&\multicolumn{2}{l}{-Compute $R_3=r_2.P$} \\
		&\multicolumn{2}{l}{-Compute $R_4= \mathbb{SK}_{CS_j}.R_3$}\\
		&\multicolumn{2}{l}{-Generate time stamp $T_{CS_j}$}\\ 
		&\multicolumn{2}{l}{-Compute $\mathbb{A}uth_{CS_j-CAG} = \mathbb{H}(R_3|| \mathfrak{PTD}_{CS_j} || \mathfrak{PTD}_{CAG} || T_{CS_j}|| \mathbb{A}uth_{EV_i-CAG})$}\\
		
		&\multicolumn{2}{c}{$\xrightarrow{M3=<\mathbb{A}uth_{CS_j-CAG},~T_{CS_j},~T_{EV_i},~r_1,~r_2,~R_1,~\mathfrak{PTD}_{EV_i},~\mathfrak{PTD}_{CS_j}>}$}\\
		
		&\multicolumn{2}{l}{\hspace*{5cm}-Validate time stamp $T_{EV_i}$ and $T_{CS_j}$}\\
		&\multicolumn{2}{l}{\hspace*{5cm}-Compute $\mathbb{A}uth_{EV_i-CAG_j}^{*}=$}\\
		&\multicolumn{2}{l}{\hspace*{5cm}$\mathbb{H}(R_1 || r_1.\mathbb{PK}_{EV_i} || \mathfrak{PTD}_{EV_i} || \mathfrak{PTD}_{CAG} || T_{EV_i} )$}\\
		
		&\multicolumn{2}{l}{\hspace*{5cm}-Compute $\mathbb{A}uth_{CS_j-CAG}^{*}= \mathbb{H}(r_2.\mathbb{PK}_{CS_j}|| \mathfrak{PTD}_{CS_j} || \mathfrak{PTD}_{CAG} || $}\\
		&\multicolumn{2}{l}{\hspace*{5cm}$ T_{CS_j}||\mathbb{A}uth_{EV_i-CAG}^*)$}\\
		&\multicolumn{2}{l}{\hspace*{5cm}-Check $\mathbb{A}uth_{CS_j-CAG}^{*} \stackrel{?}{=}  \mathbb{A}uth_{CS_j-CAG}$} \\
		&\multicolumn{2}{l}{\hspace*{5cm}-If same, $j^{th}$ CS  and $i^{th}$ EV are marked authentic; else connection}\\
		&\multicolumn{2}{l}{\hspace*{5cm}terminated}\\
		
		&\multicolumn{2}{l}{\hspace*{5cm}-Generate time stamp $T_{CAG}$}\\
		&\multicolumn{2}{l}{\hspace*{5cm}-Compute  $R_5=\mathbb{SK}_{CAG}. R_1$}\\
		&\multicolumn{2}{c}{\hspace*{5cm}-Compute $\mathbb{A}uth_{CAG}= \mathbb{H}(R_5|| \mathfrak{PTD}_{EV_i} ||\mathfrak{PTD}_{CS_j} || \mathfrak{PTD}_{CAG} || T_{CAG})$}\\
		\multicolumn{3}{c}{$\xleftarrow{M_4= <\mathbb{A}uth_{CAG},~T_{CAG}>}$} \\
		&\multicolumn{2}{l}{-Validate time stamp $T_{CAG}$}\\
		&\multicolumn{2}{l}{-Compute $\mathbb{A}uth_{CAG}^{*}= \mathbb{H}(r_1.\mathbb{PK}_{CAG}|| \mathfrak{PTD}_{EV_i} ||\mathfrak{PTD}_{CS_j} || \mathfrak{PTD}_{CAG} || T_{CAG}) $}\\
		&\multicolumn{2}{l}{-Check $\mathbb{A}uth_{CAG}^{*} \stackrel{?}{=}  \mathbb{A}uth_{CAG}$}\\
		&\multicolumn{2}{l}{-If same, CAG is marked authentic; else tear down the connection}\\
		\multicolumn{2}{c}{$\xleftarrow{M_5= <\mathbb{A}uth_{CAG},~T_{CAG},~r_1>}$}& \\
		
		\multicolumn{2}{l}{-Validate time stamp $T_{CAG}$}&\\
		\multicolumn{2}{l}{-Compute $\mathbb{A}uth_{CAG}^{*}= \mathbb{H}(r_1.\mathbb{PK}_{CAG}|| \mathfrak{PTD}_{EV_i} ||\mathfrak{PTD}_{CS_j} || \mathfrak{PTD}_{CAG} || T_{CAG}) $}&\\
		\multicolumn{2}{l}{-Check $\mathbb{A}uth_{CAG}^{*} \stackrel{?}{=}  \mathbb{A}uth_{CAG}$}&\\
		\multicolumn{2}{l}{-If same, CAG is marked authentic; else tear down the connection}&\\
		
		\hline
		
	\end{tabular}
	\caption{An illustration of the proposed mutual authentication phase based on hierarchical approach.}
	\label{fig:Phase3}
\end{figure*}

\vspace{0.7mm}
\noindent \textit{Step 1:} This phase is initiated by the $CS_j$, moment $EV_i$ connects to the charging point at the CS and begins to charge or discharge its battery for effective demand response and ancillary services. In order to initiate the process, the CS selects a random number $r_1\in Z^{*}_p$ and generates a time stamp $T_{CS_j}$. Subsequently, it computes $R_1=r_1.P$ using ECC multiplicative operation over  $r_1$ and $P$. Finally, the $j^{th}$ CS generates the first message $M_1= <R_1 || \mathfrak{PTD}_{CS_j} || T_{CS_j}>$  and relays the same to the $i^{th}$ EV for further processing.

\vspace{0.7mm}
\noindent \textit{Step 2:} On receiving $M_1$, the $i^{th}$ EV extracts the pseudo identity of the $j^{th}$ CS ($\mathfrak{PTD}_{CS_j}$) along with $R_1$ and $T_{CS_j}$. It then validates $T_{CS_j}$ if it is within the permissible time frame. Using this, the EV computes $R_2=\mathbb{SK}_{EV_i}. R_1$ using its private key $\mathbb{SK}_{EV_i}$ and received $R_1$. Next, it generates the time stamp $T_{EV_i}$ and computes two tokens for authenticating the $j^{th}$ CS ($\mathbb{A}uth_{EV_i-CS_j}= \mathbb{H}(R_1 || R_2 || \mathfrak{PTD}_{EV_i} || \mathfrak{PTD}_{CS_j} || T_{EV_i} )$) and the CAG ($\mathbb{A}uth_{EV_i-CAG}= \mathbb{H}(R_1 || R_2 || \mathfrak{PTD}_{EV_i} || \mathfrak{PTD}_{CAG} || T_{EV_i} )$), respectively.

\vspace{0.7mm}
\noindent \textit{Step 3:} Finally, the $i^{th}$ transmits the message ${M2}$ to the $j^{th}$ CS with the following tokens $<\mathbb{A}uth_{EV_i-CS_j},~\mathbb{A}uth_{EV_i-CAG},~T_{EV_i}, ~\mathfrak{PTD}_{EV_i}>$.

\vspace{0.7mm}
\noindent \textit{Step 4:} The $j^{th}$ CS initiates the process to check the authenticity of the $i^{th}$ EV as follows. Initially, its validates the time stamp $T_{EV_i}$ and proceeds only if its within the permissible time window. Next, it computes the intermediate authentication token $\mathbb{A}uth_{EV_i-CS_j}^{*}$ and compares its value with the received token $\mathbb{A}uth_{EV_i-CS_j}$. A matched value establishes the authenticity of the $i^{th}$  EV and a mismatch indicates a malicious entity leading to connection termination.	

\vspace{0.7mm}
\noindent \textit{Step 5:} In the former scenario, the CS continues with the authentication process and proceeds with generating an authentication token for the CAG as follows. Firstly, it generates a random number $r_2\in Z^{*}_p$ and then computes $R_3$ and $R_4$. Next, it generates the time stamp $T_{CS_j}$ and using the above mentioned values computes a token for CAG to authenticate the CSs $\mathbb{A}uth_{CS_j-CAG}$. Its values is equivalent to $\mathbb{H}(R_3|| \mathfrak{PTD}_{CS_j} || \mathfrak{PTD}_{CAG} || T_{CS_j}|| \mathbb{A}uth_{EV_i-CAG})$. It is worth mentioning here that this token also encapsulates the authentication token for the CAG transmitted by the $i^{th}$ EV. Finally, the CS transmits message $M3$ to the CAG with $\mathbb{A}uth_{CS_j-CAG}$,~$T_{CS_j},$~$T_{EV_i}$,~$r_1$,~$r_2$,~$R_1$,~$\mathfrak{PTD}_{EV_i}$, and~$\mathfrak{PTD}_{CS_j}$ tokens.

\vspace{0.7mm}
\noindent \textit{Step 6:} In response to the received message, the CAG initially validates the $T_{EV_i}$ and $T_{CS_j}$. It next computes the intermediate authentication tokens $\mathbb{A}uth_{CS_j-CAG}^{*}$ and $\mathbb{A}uth_{CS_j-CAG}^{*}$. Finally, it cross verifies the correctness of $\mathbb{A}uth_{CS_j-CAG}^{*}$ against $\mathbb{A}uth_{CS_j-CAG}$. Identical values, establish the authenticity of both the $i^{th}$ EV and $j^{th}$ CS, and the process proceeds further; otherwise the connection is terminated.

\vspace{0.7mm}
\noindent \textit{Step 7:} In this step, the CAG generates a token for the $i^{th}$ EV and $j^{th}$ CS, referred as $\mathbb{A}uth_{CAG}$ using time stamp $T_{CAG}$ and token $R_5=\mathbb{SK}_{CAG}. R_1$. Next it relays the message $M_4= <\mathbb{A}uth_{CAG},~T_{CAG}>$ to the CS.

\vspace{0.7mm}
\noindent \textit{Step 8:} The CS then validates the received time stamp and correctness of the authentication token $\mathbb{A}uth_{CAG}^{*}$. Valid result prove the authentication of the CAG and the received token are then transmitted to the $EV_i$ for further processing along with $r_1$. 

\vspace{0.7mm}
\noindent \textit{Step 9:} The $i^{th}$ EV repeats the above process and mutually validates the authenticity of the CAG with the received token.

\vspace{2.5mm}
\noindent \textit{\textbf{Phase 4: Consensus Mechanism}}
\vspace{1mm}

The proposed secure and anonymous energy trading mechanism employ the advantages of the PBFT consensus mechanism for maintaining the global ledger. The transaction of reward to the participating EVs  from the utility is accomplished in accordance with the following steps:

\vspace{0.7mm}
\noindent \textit{Step 1:} In the considered V2G scenario, it is assumed that the CSs are equipped with sufficient computational and communicational resources; wherein CSs have the ability to write a block to the ledger.  

\vspace{0.7mm}
\noindent \textit{Step 2:} Let us assume, a total of $n$ CSs are registered with the CAG. Amongst these CSs, one is selected as the ``Speaker" and rest are marked as ``Congressmen". The primary role of the ``Speaker"  is to organise the consensus mechanism while staying away from the voting process involved in consensus. The selected speaker is liable to conduct consensus for approximately $m$ turns. The selection of the speaker amongst the $n$ available CS candidates is based on the following rule: $x = (h ~\text{mod}~m)+1$. Here, the variables $x$ and $m$ refer to the selected speaker and height of the current block, respectively.

\vspace{0.7mm}
\noindent \textit{Step 3:} After successful authentication and availing V2G services, the $j^{th}$ CS relays the transaction details to the CAG; which then broadcast the details to all the CSs on the blockchain network. The CSs  then store the transaction details in their respective memories prior to transferring them to the ledger. 

\vspace{0.7mm}
\noindent \textit{Step 4:} Post $t$ time intervals, the block containing the transaction details is created which then undergoes the voting process carried out by the speaker. In the initial run, the speaker request congressmen to cast their votes.

\vspace{0.7mm}
\noindent \textit{Step 5:} Following this, the congressmen casts their respective votes. On the basis of the received response from congressmen, the speaker reaches a consensus to finally publish the block with the transaction details on the  global ledger.

%
\section{Results and Discussion} \label{sec:ObservationAndAnalysis}
In this section, the performance of the proposed scheme is extensively assessed in terms of different evaluation metrics such as security features support, formal security verification, and computational and communicational overhead analysis. The detailed description is mentioned as follows.

\vspace{2.5mm}
\noindent \textit{A. Security feature evaluation}
\vspace{1mm}

The proposed blockchain based hierarchical authentication mechanism supports the following security features: mutual authentication, anonymity of CSs, EVs and CAG, unforgeability, unlinkability and limited operations for the EVs. Further, it also provides replay protection with forward secrecy and prevents identity spoofing. 

\vspace{2.5mm}
\noindent \textit{B. Formal security verification}
\vspace{1mm}


In order to validate the safeness of the designed hierarchical authentication protocol (as detailed in Phase 3 of the proposed scheme), it has been subjected to an open source suite of applications named AVISPA. The tool is extensively used by the research community to validate and verify the security goals of any designed protocol against a rich source of attack vectors provided by AVISPA's back-ends namely-on the fly model checker (OFMC), CL-based attack searcher (CL-AtSe), SAT-based model checker (SATMC), and tree automata-based protocol analyzer (TA4SP).  Additionally, AVISPA is also employed to trace any security flaw in the designed protocol and devise different methods to remove it. Further, AVISPA accepts the input in the form of a role-based language known as high level protocol specification language (HLPSL). Using this language, the different entities involved in the designed protocol are expressed as different roles which interact amongst each other to trigger different transactions.
For instance, in the considered scheme, EVs, CSs and CAG were portrayed as different roles and the execution flow, as depicted in Fig.~\ref{fig:Phase3}, as different transactions. The result of executing these transactions on  OFMC and CL-AtSe back-ends lead to the ``Safe" results as shown in Fig.~\ref{fig:SPAN}. This clearly indicates that the proposed hierarchical authentication mechanism based on blockchain is safe to be executed on real-time test beds.
\begin{figure}[ht]
	\centering
	\scriptsize
	\begin{tabular}{| p{3.5cm} | p{3.5cm} |}
		\hline
		\begin{minipage}{15em}
			\scriptsize
			\ \ \\
			\% OFMC\\
			\% Version of 2006/02/13\\
			SUMMARY\\
			\hspace*{2mm}SAFE\\
			DETAILS\\
			\hspace*{2mm}BOUNDED\_NUMBER\_OF\_\\SESSIONS
			PROTOCOL\\
			\hspace*{2mm}/home/span/Hierarchical.if\\
			GOAL\\
			\hspace*{2mm}as\_specified\\
			BACKEND\\
			\hspace*{2mm}OFMC\\
			COMMENTS\\
			STATISTICS\\
			\hspace*{2mm}parseTime: 0.00s\\
			\hspace*{2mm}searchTime: 0.32s\\
			\hspace*{2mm}visitedNodes: 40 nodes\\
			\hspace*{2mm}depth: 4 plies
		\end{minipage} & 
		\begin{minipage}{15em}
			\scriptsize
			\ \ \\
			SUMMARY\\
			\hspace*{2mm}SAFE\\
			DETAILS\\
			\hspace*{2mm}BOUNDED\_NUMBER\_OF\_\\SESSIONS
			TYPED\_MODEL\\
			PROTOCOL\\
			\hspace*{2mm}/home/span/Hierarchical.if\\
			GOAL\\
			\hspace*{2mm}As Specified\\
			BACKEND\\
			\hspace*{2mm}CL-AtSe\\
			STATISTICS\\
			\hspace*{2mm}Analysed   : 0 states\\
			\hspace*{2mm}Reachable  : 0 states\\
			\hspace*{2mm}Translation: 0.24 seconds\\
			\hspace*{2mm}Computation: 0.00 seconds
		\end{minipage}\\
		\hline
	\end{tabular}
	\caption{Evaluation of mutual authentication based on hierarchical approach on AVISPA.}\label{fig:SPAN}
\end{figure}

\vspace{2.5mm}
\noindent \textit{C. Analysis of Computation and Communication Overhead}
\vspace{1mm}

\begin{figure}[!ht]
	\centering
	\includegraphics[scale=.4]{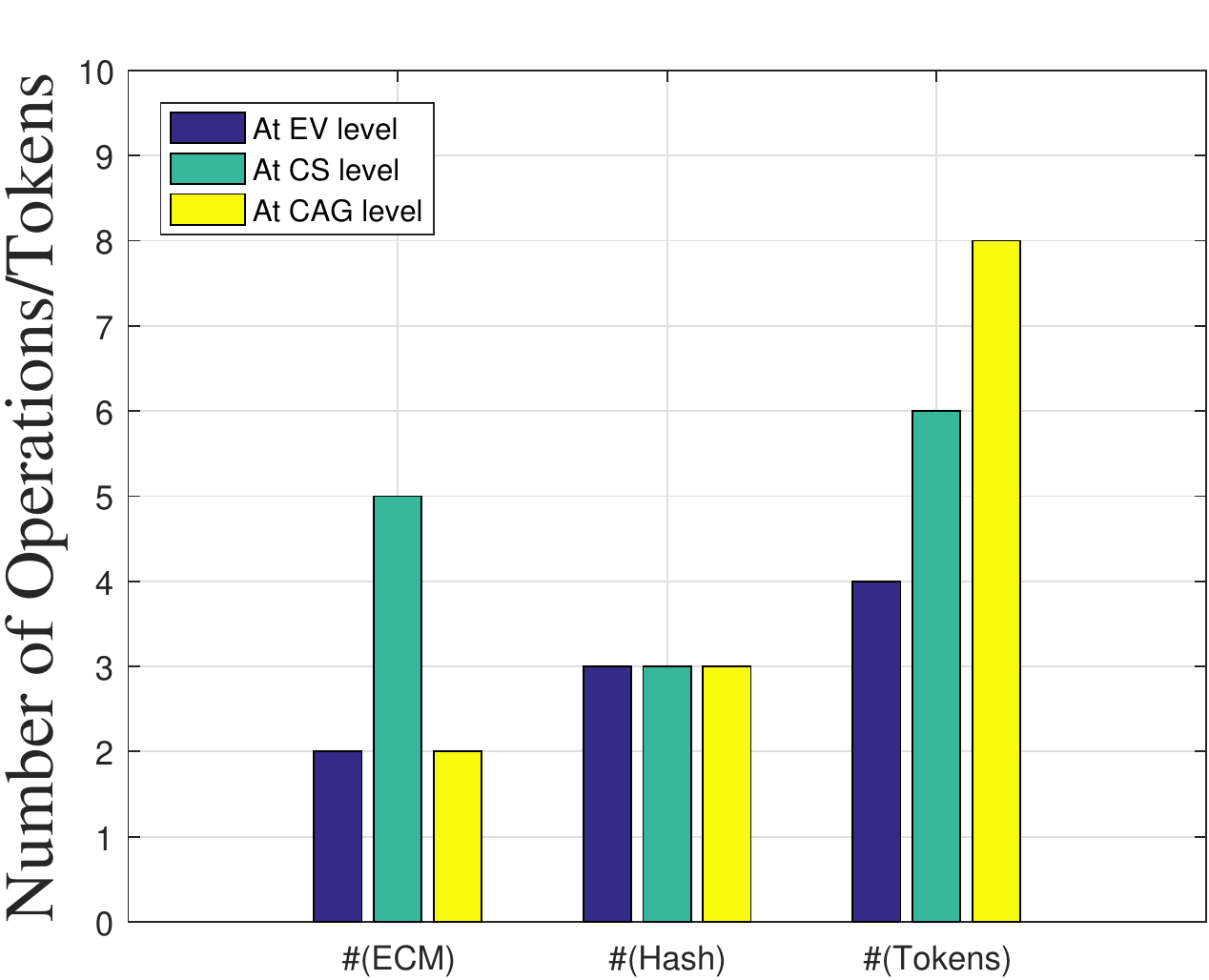}
	\caption{Overhead analysis.}
	\label{fig:fig1}
\end{figure}

In this section, we analysis the computational and communication overhead across the three entities involved in the mutual authentication process based on hierarchical approach. It is evident from the description given in Section~\ref{sec:ProposedScheme} that EVs, CSs and CAG participate in hierarchical authentication for mutually authenticating each other before the services could be availed/provided by the EVs and rewards could be granted in return. In the overall process, the considered entities incur computational and communicational expenses. Hence, this section elaborates the same in detail. The computational expenses incurred by the EVs, CSs and CAG could be attributed to the number of cryptographic operations performed in the overall process. Here, the most significant operation encompasses ECC multiplication (\textit{ECM}) operations followed by one-way hash (\textit{Hash}) operations. Fig.~\ref{fig:fig1} depicts the relative comparison between the chosen entities on the basis of number of \textit{ECM} and \textit{Hash} operations. It is evident from the obtained results that the least number of cryptographic operations are executed by the resource constrained EVs and computationally busy CAG followed by the CS. 

\noindent On the other hand, the communication overhead is expressed in terms of the number of incoming tokens. The higher the number of incoming tokens, the higher is the communicational cost. The related results have been highlighted in Fig.~\ref{fig:fig1} which clearly indicate that the EVs experience the least communication overhead followed by CS and then by CAG. Thus, it can be summarized that the designed hierarchical authentication mechanism not only guarantees enhanced security support but also imposes less overheads on the battery powered EVs.

\section{Conclusion} \label{sec:Conclusions}
With the increasing penetration of EVs in SG scenarios, distributed V2G services have witnessed a major blow in the last couple of years. Recent research statistics indicate that the efficient use of a fleet of EVs is detrimental in managing grid fluctuations. Towards this end, the need to design an efficient and secure energy trading mechanism is of utmost importance. Thus, in this paper, we proposed a blockchain based hierarchical authentication mechanism; wherein the global ledger helps in the secure and anonymous dispatch of rewards to the participating EVs. On the other hand, the proposed hierarchical mechanism helps establish the mutual authenticity of EVs, CSs and the CAG and is a novel attempt in this direction. Further, the obtained results indicate that the proposed mechanism is suitable for V2G scenarios and is apt for resource constrained EVs as it leads to reduced communicational and computational expenses.

\bibliographystyle{IEEEtran}
\bibliography{ICCRef2.bib}
\end{document}